Proceedings of

# X International Workshop on Locational Analysis and Related Problems

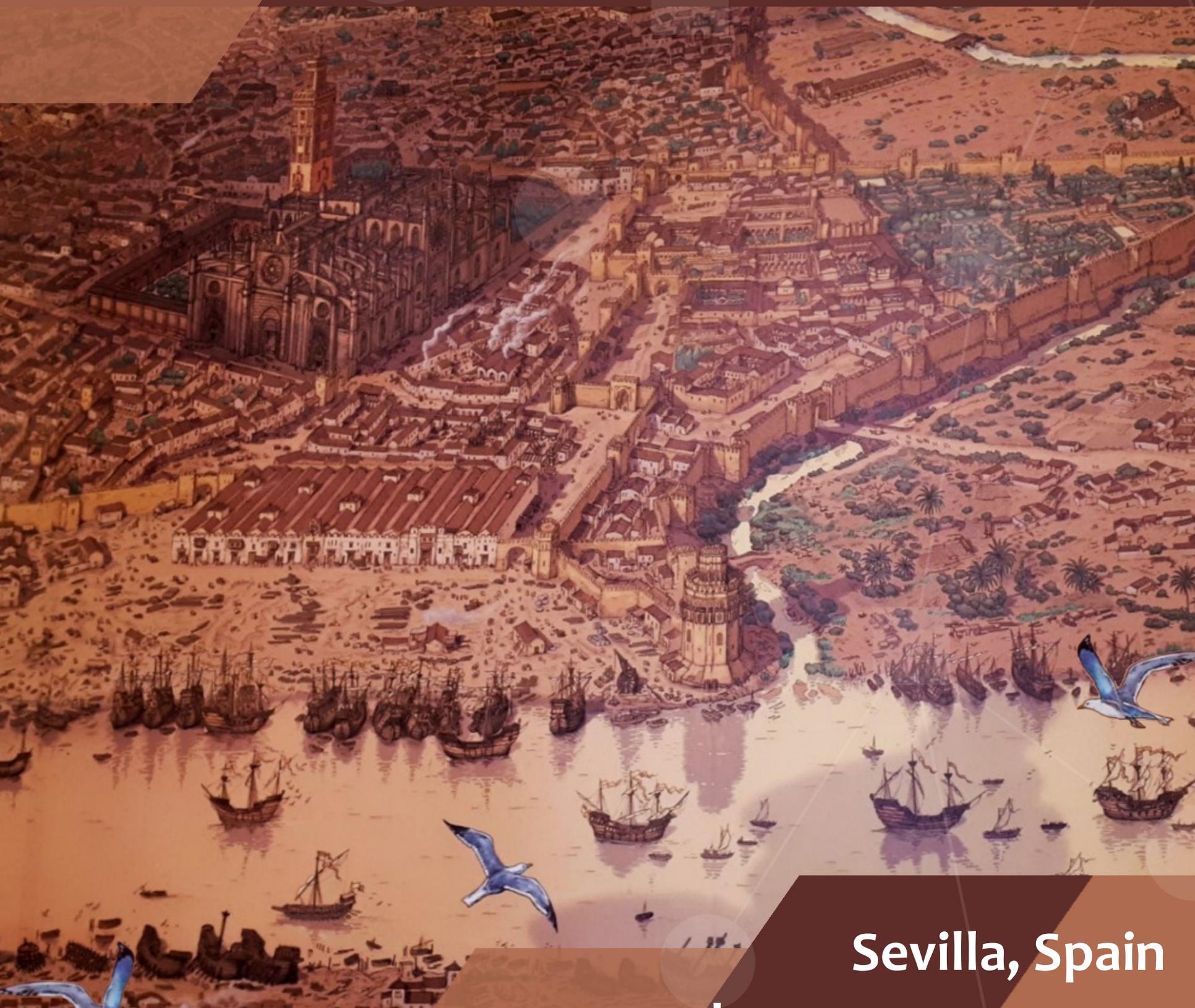

Sevilla, Spain
January 23-24, 2020

|       | Thursday Jan 23rd       | Friday Jan 24th                 |
| ----- | ----------------------- | ------------------------------- |
| 8:45  | Registration            |                                 |
| 9:15  | Opening Session         |                                 |
| 9:30  | Session 1:              | Session 5:                      |
|       | Facility Location I     | Facility Location II            |
| 11:30 | Coffee break            | Coffee break                    |
| 12:00 | Invited Speaker:        | Invited Speaker:                |
|       | Stefano Benati          | Ángel Corberán                  |
| 13:00 | Session 2:              | Session 6:                      |
|       | Applications I          | Routing I                       |
| 14:00 | Lunch                   | Lunch                           |
| 15:30 | Session 3:              | Session 7:                      |
|       | Districting             | Routing II                      |
| 16:30 | Coffee break            | Coffee break                    |
| 17:00 | Session 4:              | Session 8:                      |
|       | Logistics               | Applications II                 |
| 18:40 |                         | Location Network Meeting        |
| 19:30 | Social event            |                                 |
| 21:00 |                         | Conference dinner               |

# PROCEEDINGS OF THE X INTERNATIONAL WORKSHOP ON LOCATIONAL ANALYSIS AND RELATED PROBLEMS (2020)

# Preface

The International Workshop on Locational Analysis and Related Problems will take place during January 23-24, 2020 in Seville (Spain). It is organized by the Spanish Location Network and the Location Group GELOCA from the Spanish Society of Statistics and Operations Research(SEIO). The Spanish Location Network is a group of more than 140 researchers from several Spanish universities organized into 7 thematic groups. The Network has been funded by the Spanish Government since 2003.

One of the main activities of the Network is a yearly meeting aimed at promoting the communication among its members and between them and other researchers, and to contribute to the development of the location field and related problems. The last meetings have taken place in Cádiz (January 20-February 1, 2019), Segovia (September 27-29, 2017), Málaga (September 14-16, 2016), Barcelona (November 25-28, 2015), Sevilla (October 1-3, 2014), Torremolinos (Málaga, June 19-21, 2013), Granada (May 10-12, 2012), Las Palmas de Gran Canaria (February 2-5, 2011) and Sevilla (February 1-3, 2010).

The topics of interest are location analysis and related problems. This includes location models, networks, transportation, logistics, exact and heuristic solution methods, and computational geometry, among others.

The organizing committee.

## Scientific committee:

- Maria Albareda Sambola (Universitat Politécnica de Cataluña, Spain)
- Giuseppe Bruno (Università degli Studi di Napoli Federico II, Italy)
- Jörg Kalcsics (University of Edinburgh, United Kingdom)
- Martine Labbé (Université Libre de Bruxelles, Belgium)
- Ivana Ljubic (ESSEC, France)
- Alfredo Marín (Universidad de Murcia, Spain)
- Juan A. Mesa (Universidad de Sevilla, Spain)
- Stefan Nickel (Karlsruhe Institut für Technologie, Germany)
- Blas Pelegrín (Universidad de Murcia, Spain)
- Justo Puerto (Universidad de Sevilla, Spain)
- Antonio M. Rodríguez-Chía (Universidad de Cádiz, Spain)
- Francisco Saldanha da Gama (Universidade de Lisboa, Portugal)

## Organizing committee:

- Maria Albareda Sambola (Universitat Politècnica de Catalunya)
- Marta Baldomero Naranjo (Universidad de Cádiz)
- Yolanda Hinojosa (Universidad de Sevilla)
- Alberto Japón (Universidad Pablo de Olavide)
- Luisa Isabel Martínez Merino (Universidad de Cádiz)
- Francisco A. Ortega Riejos (Universidad de Sevilla)
- Diego Ponce (Universidad de Sevilla)
- Miguel A. Pozo (Universidad de Sevilla)
- Justo Puerto (Universidad de Sevilla)
- Victoria Rebillas Loredo (Universitat Politècnica de Catalunya)
- Moisés Rodríguez Madrena (Universidad de Sevilla)
- Dolores Rosa Santos Peñate (Universidad de Las Palmas de Gran Canaria)

# Contents

**XPP with Neighborhoods**                                                **73**
*C. Valverde and J. Puerto*

PROGRAM

# Thursday January 23rd

## 08:45-09:15 Registration

## 09:15-09:30 Opening Session

## 09:30-11:30 Session 1: Facility Location I

Extensive facility location problems with hyperplanes
*V. Blanco, A. Japón, D. Ponce, and J. Puerto*

Hub Location Problems with Ball-Shaped Neighborhoods of varying radius
*V. Blanco and J. Puerto*

Upgrading Location Problems with Edge Length Variations on Networks
*M. Baldomero-Naranjo, J. Kalcsics, and A.M. Rodríguez-Chía*

Solving Discrete Ordered Median Problems with Induced Order
*E. Domínguez and A. Marín*

Upgraded network $p$-median problem
*I. Espejo and A. Marín*

## 11:30-12:00 Coffee break

## 12:00-12:50 Invited Speaker: Stefano Benati

Clustering with Variables Selection: Challenges from the Eurobarometer Data

## 13:00-14:00 Session 2: Applications I

Multi-channel distribution in the banking sector and network restructuring
*S. Baldassarre, G.Bruno C. Piccolo, and D. Ruiz-Hernández*

A Facility Location Problem with transhipment points
*A. Moya-Martínez, M. Landete, and J.F. Monge*

A Location-Allocation Model for Bio-Waste Management
*D.R. Santos-Peñate, R.R. Suárez-Vega, and C. Florido de la Nuez*

## 14:00-15:30 Lunch

## 15:30-16:30 Session 3: Districting

Shape and Balance in Districting
*V. Bucarey and F. Ordóñez*

Optimal Site Dimensioning in Design of UAV-based 5G Networks with Optical Rings, Solar Panels and Batteries.
*L. Amorosi*

Towards a stochastic programming modeling framework for districting
*A. Diglio, S. Nickel, and F. Saldanha da Gama*

## 16:30-17:00 Coffee break

## 17:00-18:40 Session 4: Logistics

The facility location problem with capacity transfers
*Á. Corberán, M. Landete, J. Peiró, and F. Saldanha-da-Gama*

A two-echelon distribution network under customer selection
*H.I. Calvete, C. Galé, and J.A. Iranzo*

Promoting the selective collection of urban solid waste by means an optimal deployment of routes for mobile eco-points
*J.A. Mesa, F.A. Ortega, and R. Piedra-de-la-Cuadra*

Dynamic shipping in hierarchical freight transportation network
*I. Contreras, G. Laporte, and D. Ponce*

## 19:30 Social event

# Friday January 24th

## 09:30-11:30 Session 5: Facility Location II

Minimum distance regulation and entry deterrence through location decisions
*J. Elizalde and I. Rodríguez-Carreño*

Facility Location Problems on Multilayer Networks
*M. Calvo and J.A. Mesa*

Infrastructure Network Design Models: some previous facts before Benders Decomposition
*V. Bucarey, B. Fortz, M. Labbé, N. González-Blanco, and J.A. Mesa*

An application of the $p$-median problem in optimal portfolio selection
*J. Puerto, M. Rodríguez-Madrena, and A. Scozzari*

$p$-Maximum expected covering on an unreliable network
*M. Albareda-Sambola and O. Lordan*

## 11:30-12:00 Coffee break

## 12:00-12:50 Invited Speaker: Ángel Corberán

Arc Routing Problems with drones

## 13:00-14:00 Session 6: Routing I

Distance-Constrained Close Enough Arc Routing Problem: Polyhedral study and B&C algorithm
*Á. Corberán, I. Plana, M. Reula, and J. M. Sanchis*

A heuristic approach for the multi-drop Truck-Drone cooperative routing problem
*P.L. González-R, D. Canca, J.L. Andrade-Pineda, M. Calle, and J.M. León*

Orienteering with synchronization in a telescope scheduling problem
*J. Riera-Ledesma, J.J. Salazar-González, and F. Garzón*

**14:00-15:30 Lunch**

**15:30-16:30 Session 7: Routing II**

A matheuristic to solve the MDVRP with Vehicle Interchanges
*V. Rebillas-Loredo, M. Albareda-Sambola, J.A. Díaz, and D.E. Luna-Reyes*

Prize-collecting Location Routing on Trees
*J. Aráoz, E. Fernández, and M. Munoz-Marquez*

Network Design under Uncertain Demand: an Alternative Capacitated Location Decision Framework
*D. Ruiz-Hernández, M.M.B.C Menezes, and O. Allal-Cherif*

**16:30-17:00 Coffee break**

**17:00-18:40 Session 4: Applications II**

An approach to obtain the Domatic Number of a graph by block decomposition
*M. Landete and J.L. Sainz-Pardo*

XPP with Neighborhoods
*C. Valverde and J. Puerto*

Optimisation Models Applied to Image Reconstruction
*J.J. Calvino, M. López-Haro, J. M. Muñoz-Ocaña, and A.M. Rodríguez-Chía*

On some classes of combinatorial optimization problems with non-convex neighborhoods
*I. Espejo, J. Puerto, and A.M. Rodríguez-Chía*

**18:40-19:10 Location Network Meeting**

**21:00 Conference dinner**

INVITED SPEAKERS

# Arc Routing Problems with drones

Ángel Corberán[1]

[1]*Dept. d'Estadística i Investigació Operativa, Universitat de València, Spain,*
angel.corberan@uv.es

In this talk we present some drone arc routing problems (Drone ARPs) and comment their relation with well-known postman ARPs. Applications for Drone ARPs include traffic monitoring, infrastructure inspection, and surveillance along linear features such as coastlines or territorial borders. Unlike the postmen in traditional ARPs, drones can travel directly between any two points in the plane without following the edges of the network. Hence, a drone route may service only part of an edge, with multiple routes being used to cover the entire edge. Thus the Drone ARPs are continuous optimization problems in which the shape of the lines to service has to be considered. To deal with this feature we discretize the problems by approximating each curve by a polygonal chain and allowing the drones to enter and leave each curve only at the points of the polygonal chain. If the capacity of the vehicles is unlimited, the resulting problem is a rural postman problem (RPP), otherwise we have the Length Constrained $K$-Drones Rural Postman Problem (LC K-DRPP). The LC K-DRPP is an ARP where there is a fleet of homogeneous drones that have to service a set of edges of a network. The drones routes have to start and end at a given vertex and, since the autonomy of the drones is restricted, the length of their routes is limited by a maximum distance. For this problem, we propose a matheuristic algorithm and present an ILP formulation and a polyhedral study of its solutions. A preliminary branch-and-cut algorithm is also introduced and some computational results are presented.

This is a joint work with **James F. Campbell**, **Isaac Plana**, **José M. Sanchis**, and **Paula Segura**

# Clustering with Variables Selection: Challenges from the Eurobarometer Data

Stefano Benati[1]

[1]*School of International Studies, Faculty of Sociology and Social Research University of Trento, Italy*   stefano.benati@unitn.it

Clustering algorithms may fail to find good units partitions because they run on input containing too many variables. When the data set is large, it can contain hundredths of variables of interest, but only a few of them are important to separate groups. Useless variables are called "masking", but their inclusion into the model results in an imprecise distance/dissimilarity measure. We will describe a model, based on the $p$-median or on the $k$-means algorithms for clustering, that includes the variable/feature selection among the decision variables. The problem is at least as complex as the original classification algorithm, but thanks to its Integer Linear Programming (ILP) formulation, it can be solved in an exact way at least for moderate instances size. Moreover, the ILP formulation is useful to formulate heuristic procedures for the instances that are not solved exactly.

We show the relevance of these models applying our algorithms to Eurobarometer data. These surveys record the opinions of the European citizens and they are an important source of knowledge for social and political scientists. Yet, the way in which questions are formulated (multiple items answers, Likert scales and so on), are posing important challenges to data scientists, as those data do not fit into the standard assumptions of ordinary inference (multivariate normality, urn sampling, and so on). We will show how the feature selection clustering facilitates the interpretation of the output, while, at the same time, the application is suggesting new further and original development for clustering models.

ABSTRACTS

# $p$-Maximum expected covering on an unreliable network[*]

## Maria Albareda-Sambola[1] and Oriol Lordan[2]

[1]*Statistics and Operations Research department, Universitat Politècnica de Catalunya (UPC)*,  maria.albareda@upc.edu

[2]*Management department, Universitat Politècnica de Catalunya (UPC)*,  oriol.lordan@upc.edu

This work addresses the problem of locating $p$ facilities in a graph to provide an essential service to the nodes assuming that edges are prone to failure. With the aim of optimizing the accessibility to this service, the goal is to maximize the expected number of nodes that remain connected to at least one of the located facilities after failures occur. We propose a heuristic method for this problem that is based on an approximation on trees.

## The problem

Reliability of stochastic networks has motivated several lines of research from different perspectives such as computer science, graph theory, or operations research. These include, among others, defining measures of network reliability, developing algorithms for computing them efficiently or designing networks with high values of these measures (see, for instance [1] or [4]). In this work we consider a discrete location problem defined on a network with unreliable edges.

Formally, we are given an integer $p \geq 1$ and an undirected network $N = (V, E)$ whose edges might fail. The joint probability distribution of these failures is also given. Each possible scenario $\omega \in \Omega$ under this joint probability distribution is characterized by its probability, $p^w$, and the set of edges that remain available after the failures occur: $E^\omega \subseteq E$. Given such a scenario and a pair of nodes $u, v \in V$, we will say that $u$ is covered by $v$

[*]This research was partially funded by the Spanish Ministry of Economy and Competitiveness and ERDF funds through Grant MTM2015-63779-R

under scenario $w$ ($u \in N^w(v)$) if there exists an $u$–$v$ path in $G$ using only edges from $E^w$. The goal of the p-Maximum Expected Cover ($p$MEC) is to determine $p$ nodes of the network where to locate some facilities, so that the expected number of nodes covered by at least one of the facilities is maximized:

$$C^* \in \arg\max \left\{ \mathbf{E} \left| \bigcup_{v \in C} N^w(v) \right| \text{ with } C \subseteq V, |C| = p \right\}$$

This problem has been addressed previously in the literature, most often restricted to situations where only one edge/node can fail ( [2]) or to very particular assumptions on the probability distribution of edge failures ( [3]). In this work, like in [5], we assume that edge failures take place independently of each other. For this case, we propose a MIP formulation that provides good bounds on trees. Based on this formulation, we develop a heuristic method for general networks. In this last case, Montecarlo simulation is used to evaluate the candidate solutions.

# Optimal Site Dimensioning in Design of UAV-based 5G Networks with Optical Rings, Solar Panels and Batteries

Lavinia Amorosi

*Department of Statistical Sciences, Sapienza University of Rome*
lavinia.amorosi@uniroma1.it

This talk faces the problem of designing a 5G network architecture to provide internet coverage. Indeed, thanks to the continuous development of this emerging technology, many different services can be effectively supported by the adoption of Unmanned Aerial Vehicles (UAVs) (see [3] for a survey on UAVs different civil usages). The proposed architecture is composed of 5G Base Stations carried by Unmanned Aerial Vehicles (UAVs) and supported by ground sites interconnected through optical fiber links. We also consider the dimensioning of each site in terms of the number of Solar Panels (SPs) and batteries. We then formulate the problem of cost minimization of the aforementioned architecture, by considering: i) the cost for installing the sites, ii) the costs for installing the SPs and the batteries in each site, iii) the costs for installing the optical fiber links between the installed sites, and iv) the scheduling of the UAVs to serve all areas. Our results, obtained over realistic scenarios and presented in [1] and [2], reveal that the proposed solution is effective in limiting the total costs, while being able to ensure the coverage overall areas.

This talk summarizes part of the joint work with Chiaraviglio, L., Blefari-Melazzi, N., Dell'Olmo, P., Lo Mastro, A., Natalino, C., Monti, P.

# Prize-collecting Location Routing on Trees

Julián Aráoz[1], Elena Fernández[2], and Manuel Munoz-Marquez[2]

[1]*Department of Statistics and Operation Research, Universitat Politècnica de Catalunya-BcnTech, Spain*   julian.araoz@upc.edu

[3]*Department of Statistics and Operation Research, Universidad de Cádiz, Spain*
elena.fernandez@uca.es    manuel.munoz@uca.es

In this work we study Prize-collecting Location Routing Problems (PLRPs) on trees. In PLRPs there is a set of users with demand, located at the vertices of a tree, each of which is associaed with a profit, and each edge has a cost. To serve the demand a set of routes must be set up, each of them starting and ending at a node.

The objective is to maximize the total net profit, defined as the total income from service to the selected demand vertices, minus the total cost that includes the overall set-up cost of activated facilities plus the routing cost of the edges used in the routes.

A mathematical programming formulation is presented, which has the integrality property. The formulation models a directed forest where each connected component hosts at least one open facility, which becomes the root of the component. The problems considered can also be optimally solved with ah-hoc solution algorithms.

Optimality conditions are developed, which can be exploited algorithmically in a preprocess phase and reduce substantially the size of the initial graph.

# Multi-channel distribution in the banking sector and network restructuring

Silvia Baldassarre[1], Giuseppe Bruno[1], Carmela Piccolo[1], and Diego Ruiz-Hernández[2]

[1]*Università degli studi di Napoli Federico II, Department of Industrial Engineering, Piazzale Tecchio 80, 80125, Napoli Italy*
silviabaldassarre93@libero.it; giuseppe.bruno@unina.it; carmela.piccolo@unina.it

[2]*Sheffield University Management School, Conduit Road, S10 1FL, Sheffield, UK*
d.ruiz-hernandez@sheffield.ac.uk

Banking groups in Europe are facing the challenge of restructuring their branch networks, which are typically oversized due to the market overlapping strategies prevailing during the late 1990s and early 2000s.

Additionally, the digital transformation has had a disruptive effect on the ways firms interact with their customers. As a result, companies that traditionally relied on "brick-and-mortar" facilities, aiming at a face-to-face interaction with their clients, are adopting online distribution channels to further expand their customer bases and to remain competitive.

In the banking sector, the technological innovations that led first to the diffusion of the phone banking and the Automated-Teller Machines (ATMs), and later of the online and mobile banking, put in crisis the traditional *"bricks-and-mortar"* banking model. As a consequence, banking groups began to re-think their business models and to evolve toward *"click-and-mortar"* systems, in which digital channels are combined with the classical ones. In practical terms, they seek to restructure their networks by both closing redundant branches and redefining the role of the remaining ones.

In this work, we analyse the case of the Italian banking group (Intesa Sanpaolo), which counts about four thousand branches over the counntry and 11.8 million of customers. Recently, it engaged in a significant restructuring process of their network by defining the following main strategies:

- **Branch Reduction**. Eliminating redundant or superfluous branches.

- **Branch specialization**. Adapting branches to different market segments. It involves the transition from the traditional model to a *multiformat system*, where different typologies of branches, providing services at different levels and customers, are present.

- **Externalization of counter transactions**. Involving tobacconists as external actors that supplement branches in the provision of basic financial transactions. The spirit is to guarantee proximity to the customers whilst mantaining an adequate presence over the territory.

In facility location literature, there is an emerging research stream dealing with the territorial reorganization of facility networks [1] and specific applications to the banking sectors have been also investigated [2,3].

The aim of this work is to analyse Intesa Sanpaolo's problem within the location theory framework and to provide a mathematical programming model to support the restructuring process. We propose a hierarchical covering model that determines the branches to be closed and the type of the ones that remain open. The objective is to serve all the demand for banking services and to minimize the market share ceded to competitors.

Three types of facilities are considered, each with an associated covering radius. The radii represent different accessibility conditions to be guaranteed to the users. With the aim of minimizing the market share captured by the competitors, we set a *vicinity constraint*, by defining a minimal distance between any open branch and a competitor's facility.

The model has been tested on the banks network in the city of Naples. The obtained results show the capability of the model to provide interesting scenarios and fruitful managerial implications.

# Upgrading Location Problems with Edge Length Variations on Networks[*]

Marta Baldomero-Naranjo[1], Jörg Kalcsics[2], and Antonio M. Rodríguez-Chía[1]

[1]*Departamento de Estadística e Investigación Operativa, Universidad de Cádiz, Spain,* marta.baldomero@uca.es    antonio.rodriguezchia@uca.es

[2]*School of Mathematics, University of Edinburgh, United Kingdom,*  joerg.kalcsics@ed.ac.uk

In this talk, we focus our research on the upgrading version of the maximal covering location problem with edge length variations on networks, see e.g. [1, 2] for other upgrading versions of location problems.

Let $G = (V, E)$ be an undirected network with node set $V$, edge set $E$, and non-negative node weights. For each edge, we are given its current length and an upper bound on the maximal reduction of its length. Moreover, we are given the cost per unit of reduction for each edge, which can be different for each edge, and a total budget for reductions.

The upgrading maximal covering location problem with edge length modifications aims at locating $p$ facilities to maximize the coverage taking into account that the length of the edges can be decreased subject to a budget constraint. Therefore, we look for both solutions: the optimal location of $p$ facilities and the optimal upgrading network.

In this paper, we propose algorithms whose complexity is polynomial or pseudo-polynomial to solve the problem in some particular networks as paths and trees. In addition, we propose a mixed-integer formulation of the problem for general graphs. Finally, we develop some strategies including valid inequalities and preprocessing for making the formulation solvable in a shorter time. The performance of the proposed resolution method will be tested on a set of networks.

[*]Thanks to the support of Agencia Estatal de Investigación (AEI) and the European Regional Development's funds (FEDER): project MTM2016-74983-C2-2-R, Universidad de Cádiz: PhD grant UCA/REC01VI/2017, Telefónica and the BritishSpanish Society Grant.

# Extensive facility location problems with hyperplanes

Víctor Blanco[1], Alberto Japón[2], Diego Ponce[2], and Justo Puerto[2]

[1]*IEMath-Granada, Universidad de Granada, Granada, Spain* vblanco@ugr.es

[2]*IMUS, Universidad de Sevilla, Sevilla, Spain* ajapon1@us.es dponce@us.es puerto@us.es

The determination of a hyperplane fitting a set of points is a classical problem which has been addressed in different fields. From the Location Analysis point of view this problem generalizes the classical (continuous) point facility location problem, stated by Weber [1]. A set of demand point (customers) on the plane formulates and motivates the interest of finding the position of a new facility which minimizes the weighted sum of the Euclidean distances from the facility to the users. On the other hand, in Statistics and Data Analysis, the construction of a hyperplane whose sum of squared vertical distance is minimum is a crucial step in Linear Regression using the Least Sum of Squares method (Gauss, [2]). Apart from the classical problems, one can find recent useful applications, both in Location Science and Data Analysis, of finding optimal hyperplanes fitting a set of points. Also in Classification, the widely used Support Vector Machine methodolgy [3], is based on constructing a hyperplane fitting a set of points.

In this work we study the problem of locating a given number of hyperplanes minimizing a globalizing function of the closest distances from points to hyperplanes. We propose a general framework for the problem in which general norm-based residuals and ordered median aggregations of the residuals are considered. A compact formulation is presented for the problem and also a set partitioning-based formulation is derived. The set partitioning formulation is analyzed and a Column Generation procedure is proposed for solving the problem by adequately performing preprocesing, pricing and branching. A matheuristic algorithm is also derived from the set partitioning formulation. Finally the results of an extensive computational experience are reported.

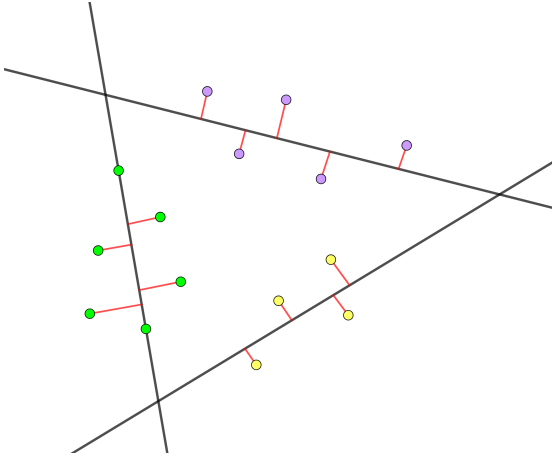

*Figure 1.*　3 hyperplanes fitting a set of points

# Hub Location Problems with Ball-Shaped Neighborhoods of varying radius

Víctor Blanco[1] and Justo Puerto[2]

[1]*IEMath-GR, Universidad de Granada, Spain*   vblanco@ugr.es

[2]*IMUS, Universidad de Sevilla, Spain*   puerto@us.es

In this work we propose an extension of the Uncapacitated Hub Location Problem where the potential positions of the hubs are not fixed in advance. Instead, they are allowed to belong to a region around an initial discrete set of nodes. We give a general framework in which the collection, transportation and distribution costs are based on norm-based distances and the hub-activation set-up costs depend, not only on the location of the hubs that are opened but also on the size of the regions where they are positioned. Two alternative mathematical programming formulations are proposed for the problem. The results of an extensive computational experience are reported showing the advantages of each of the approaches.

## 1.    Introduction

Hub-and-spoke networks have attracted the attention of the Locational Analysis community in the recent years since they allow to efficiently route commodities between customers in many transportation systems. In these networks, the flow between customers, instead of being sent directly user-to-user, is routed via a few transhipment points, *the hub nodes*. In this way, the overall transportation costs are reduced because of the economies of scale induced when sending a large amount of flow through the hub arcs. Hub Location Problems (HLP) combine the determination of the optimal placement of the hub facilities with the best routing strategies in the induced hub-and-spoke network.

   In discrete Hub Location problems one considers a given network, and the decision on locating the hubs is restricted to the positions of the given nodes. In case the network represents spatial nodes and arcs, one may obtain better results if some flexibility is allowed for the location of the hubs in the given space where the nodes of the network are embedded. Such an issue can be modeled via *neighborhoods* in which each selected node is allowed not only to be positioned in its exact original coordinates, but in a region around it, its neighborhood. This approach is particularly useful in telecommunication networks in which the decision-maker provides preferred regions in which the nodes want to be located, instead of providing a precise set of positions for them [1,2,4]. However, as far as we know, there is no previous research on the simultaneous determination of the optimal size of the neighborhoods. Observe that the neighborhoods represent locational imprecision or flexibility on the placement of objects under analysis. Thus, fixing a prespecified neighborhood size implies that better solutions are not contemplated by slightly modifyng the neighborhood size, so it is particurlarly convenient in the hub-and-spoke networks under analysis.

We introduce here an extension of the classical uncapacitated single-allocation HLP, that we call the HLP with Neighborhoods (HLPN). In this problem we are given a set of demand points, a set of potential hubs (as coordinates in $\mathbb{R}^d$), an OD flow matrix between demand points, a set-up cost for opening each potential hub, a cost measure in $\mathbb{R}^d$ and for each potential facility, a neighborhood shape which represent some flexibility on the decision of the location of the hub, and a cost based on the size of the neighborhood. The goals of HLPN are: to decide how many and which hubs must be open and which demand points are assigned to which hub; to choose the size of the neighborhood for each of the open hubs; and to find the location of the hub-servers in the neighborhood for each of its assigned demand point. All of this at minimum global cost (transportation, collection, distribution and set-up costs).

# Infrastructure Network Design Models: some previous facts before Benders Decomposition

Víctor Bucarey[1], Bernard Fortz[1], Martine Labbé[1], Natividad González-Blanco[2], and Juan A. Mesa[2]

[1]*Université Libre de Bruxelles, Brussels, Belgium*

[2]*University of Seville, Seville, Spain*

In recent years, there exist some reasons why new rail transit systems have been constructed or expanded in determined agglomerations, or are being planned for construction. The design of rapid transit networks is a powerful tool for politicians to influence users behavior and drive the communities towards a sustainable mobility. Therefore, methodological contributions that allow to improve these network designs have a true potential to have a beneficial impact in the society. The main value of this particular work are the advances it provides towards such a methodological contribution. The *Rapid Transit Network Design Problem* consists of locating alignments and stations covering as much as possible, or saving as much as possible and covering a percentage, knowing that the demand makes its own decisions about the transportation mode.

The infrastructure design problem has been treated in some papers as in [1].We propose some modifications of the model in order to improve the computational time for medium size networks. It is assumed that the mobility patterns in a metropolitan area are known and as well as, locations of potential stations and links. In addition, there already exists a different mode of transportation competing with the railway to be built.

At present, these models are difficult to solve because they have a lot of binary variables and, of course, constraints. Branch&Bound does not get to solve them in a efficient way. This reason has motivated us to explore the structure of the model and the data in order to design an efficient implementation of the Benders' Decomposition algorithm.

# Shape and Balance in Districting

Victor Bucarey[1] and Fernando Ordóñez[2]

[1]*Université Libre de Bruxelles, Brussels Belgiulm* vbucarey@ulb.ac.be

[2] *Universidad de Chile, Santiago, Chile,* fordon@dii.uchile.cl

The problem of police districting consists in dividing a geographical region in subregions (quadrants) under several considerations such as balance in the demand of police resources, geographical contiguity and compactness (see [3], [2]). In this work, a Mixed Integer Linear Programming Model is considered for designing the shape patterns of the blocks of PCSP (*Plan Cuadrante de Seguridad Preventiva*) for the national police in Chile.

This model takes into account constraints of loading resemblance according to the police resources demands and geometry guidelines. The shape considerations are related with the moment of inertia and the size of the border of each district. These considerations makes this problem untractable for realistic instances which were solved by an implementation of the Location Allocation Heuristic. This work is an extension of [1].

# A two-echelon distribution network under customer selection [*]

Herminia I. Calvete[1], Carmen Galé[2], and José A. Iranzo[3]

[1]*Statistical Methods Department, IUMA, University of Zaragoza, Pedro Cerbuna 12, 50009 Zaragoza, Spain*,  herminia@unizar.es

[2]*Statistical Methods Department, IUMA, University of Zaragoza, María de Luna 3, 50018 Zaragoza, Spain*,  cgale@unizar.es

[3]*Centro Universitario de la Defensa de Zaragoza, IUMA, Carretera de Huesca s/n, 50018 Zaragoza, Spain*,  joseani@unizar.es

Logistics operators face a very complex problem when aiming to determine how to distribute commodities to final customers to meet their expectations while remaining cost-effective. Multi-echelon systems which include two or more levels of distribution provide an appropriate structure for last-mile delivery. In this research, we consider a two-echelon distribution network consisting of a central warehouse, a set of depots and a set of customers.

The goods to be distributed to customers are stored in the central warehouse managed by the logistic operator who establishes the delivery routes of a fleet of homogeneous vehicles available at the central warehouse. The logistic operator also decides on the locations which are visited by a route. These locations can be either depots or customers. After knowing the locations which will be visited, customers are allowed to select their most convenient available location to travel to pick up their goods.

This work addresses the use of bilevel programming models to deal with the logistic operator's optimization problem which include in the constraint set the selection of the customers. Different strategies will be considered, from the point of view of the logistic operator, to take into account different goals such as minimizing the total cost of serving the routes or maximizing the quality of service to customers. First, we present the mathematical formulation of the models which can involve one or more objectives. Then, an algorithm is developed within the general framework of evolutionary algorithms that allows us to deal with those models. A set of test problems will be used to show the effectiveness of the algorithm as well as to illustrate the differences among the strategies.

---

[*]This research has been funded by the Spanish Ministry of Economy, Industry and Competitiveness under grants ECO2016-76567-C4-3-R and by the Gobierno de Aragón under grant E58 (FSE) and E41-17R (FEDER 2014-2020 "Construyendo Europa desde Aragón").

# Optimisation Models Applied to Image Reconstruction

Jose J. Calvino[2], Miguel López-Haro[2], Juan M. Muñoz-Ocaña[1], and Antonio M. Rodríguez-Chía[1]

[1]*Departamento de Estadística e Investigación Operativa.*
*Universidad de Cádiz*   juanmanuel.munoz@uca.es, antonio.rodriguezchia@uca.es

[2]*Departamento de Ciencia de los Materiales e Ingeniería Metalúrgica y Química Inorgánica.*
*Universidad de Cádiz*   jose.calvino@uca.es, miguel.lopezharo@uca.es

Electron tomography is a technique for imaging three-dimensional structures for the analysis of inorganic materials. This technique consists in reconstructing objects from a series of projections acquired thanks to a microscope. In conventional Electron Tomography, a parallel beam of electrons goes through a sample to change of their intensity. This information about the intensity of the electron beam, which is called sinogram, is recorded in a special surface. Sinograms are the inputs of our reconstruction models.

The reconstructed object is assumed as an unknown array $x$. According to the sinogram $b$ , a set of algebraic equations can be formulated as $Ax = b$, where $A$ is a matrix which contains information about every projection. The elements of the weighting matrix $A$ represent the contribution of a specific pixel to each projection beam.

Total Variation Minimization (TVM) is one of the most important reconstruction models. This model is based on minimizing the image gradient and the difference between the original sinogram ($b$) and the reconstructed one ($Ax$),

$$\min_x \sum_i ||D_i x||_2 + ||Ax - b||_2,$$

where $D$ is a $n^2 \times n^2$ matrix which calculates the image gradient and $n^2$ is the total number of pixels. Matrix $D$ is written as follows

$$D = \begin{bmatrix} 1 & -1 & 0 & ... & 0 \\ 0 & 1 & -1 & ... & 0 \\ & & & ... & \\ 0 & ... & 0 & 1 & -1 \\ -1 & 0 & ... & 0 & 1 \end{bmatrix}.$$

This model is a quadratic minimization problem, which is solved thanks to an augmented Lagrangian algorithm. We study some lineal models to obtain good quality reconstructions and some techniques to improve the computational time due to the large number of constraints. Moreover, choosing a specific group of projections plays an important role because of the fact that a good selection one will provide a high image quality. We will consider that a good projection set contains as much information as possible from the original image. This information will be calculated by maximizing the difference between every sinogram value $b_\theta^k$ and the following one $b_\theta^{k+1}$ along each projection $\theta$.

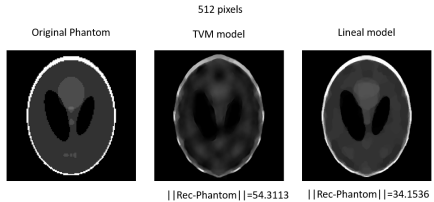

*Figure 1.* Quality difference between a TVM model and a lineal model.

# Facility Location Problems on Multilayer Networks [*]

María Calvo[1] and Juan A. Mesa,[2]

[1]*Universidad de Sevilla, Spain,*  mariacalvgonz@gmail.com

[2]*Dpto. de Matemática Aplicada II, Escuela Técnica Superior de Ingeniería, Universidad de Sevilla, Spain*  jmesa@us.es

During last decades an interdisciplinary research effort has been dedicated to modeling complex systems that cannot be represented within the classical network theory. The main characteristics of complex systems cannot be captured by a single network, or even by the projection of several networks, but by a multilayer network consisting of several networks with different features, and some connections between layers. This is for example the case of the transit networks in big cities or metropolitan areas where several modes of transportation (metro, bus, commuter trains, etc.) coexist but each mode has its proper structure and properties so that they should be represented in different layers. Another example is the system of all connections of several airline companies. Since often a trip consists of two branches with a stopover between them, the setting for mathematical problems has to be a multilayer network. Finally, a multi-storey building with different floor ocupations should again be represented by a multilayer network.

A multilayer network is a pair ,$\mathcal{M} = (\mathcal{G}, \mathcal{C})$ where

$\mathcal{G} = \{G_k, k \in \{1, 2, \ldots, K\}\}$ is a family of (networks) graphs $G_k = (V_k, E_k)$

and $\mathcal{C} = \{E_{kl} \subset V_k \times V_l; \ k, l \in \{1, 2, \ldots, K\}, k \neq l\}$
is the set of connections between vertices of different layers. Thus we have intralayer (arcs) edges and interlayer (arcs) edges.

Some faciilty location problems on multilayer networks can be solved by considering the projection network $proj(\mathcal{M}) = (V_\mathcal{M}, \mathcal{C}_\mathcal{M})$ of the multilayer network where

$$V_{\mathcal{M}} = \cup_k V_k, \text{ and } C_{\mathcal{M}} = (\cup_{j=1}^K E_k) \cup_{j,k=1, j \neq k}^K E_{kl}.$$

---
[*]This research is partially supported by project MTM2015-67706-P (MINECO/FEDER,UE).

This representation is not useful for some problems because it assumes the layer networks are similar, and the structure of a multilayer network fits better to those problem with heterogeneous networks.

In this paper, we study the extension to this framework of the classical location problems: the median, the center and the covering.

# Dynamic shipping in hierarchical freight transportation network[*]

Ivan Contreras[1,2], Gilbert Laporte[2,3], and Diego Ponce[2,4]

[1]*Concordia University, Montréal, Canada*,   icontrer@encs.concordia.ca

[2]*CIRRELT, Montréal, Canada*,   gilbert.laporte@cirrelt.net

[3]*HEC Montréal, Montréal, Canada*

[4]*IMUS and Departamento de Economía Aplicada I, Universidad de Sevilla, Sevilla, Spain*, dponce@us.es

We introduce a new problem which lets us model the situation when a retailer can enter the supply chain of a third-party logistics provider (3PL) at various levels. According to [4] and to [3], there is a significant advantage to be gained by using a 3PL to coordinate a supply chain. The static version of the problem is modeled as an integer linear program after the definition of the appropriate decision graph.

Later we use the benchmark given by this static formulation to compare four policies thought to solve the dynamic version of the problem. The basic problem underlying this second part of our study is the dynamic multi-period routing problem [1,2,6,7]. For the dynamic problem, the orders are gradually revealed over a rolling planning horizon and the proposed policies let us take decisions for every period based on rules (one algorithm) or on the solution of an integer linear program (three policies optimize the cost by solving a mathematical model).

This work shows the usefulness of the proposed policies according with extensive computational results based on real data. In addition, some interesting results about consolidation [5] are reported.

---

[*]This work has been partly funded by the Horizon Postdoctoral Fellowship, Concordia University and by the Canadian Natural Sciences and Engineering research Council under grants 2018-06704 and 2015-06189. This support is gratefully acknowledged.

# The facility location problem with capacity transfers

Ángel Corberán[1], Mercedes Landete[2], Juanjo Peiró[1], and Francisco Saldanha-da-Gama[3]

[1]*Departament d'Estadística i Investigació Operativa. Facultat de Ciències Matemàtiques, Universitat de València, Spain,*  angel.corberan@uv.es, juanjo.peiro@uv.es

[2]*Departamento de Estadística, Matemáticas e Informática. Instituto Centro de Investigación Operativa. Universidad Miguel Hernández de Elche, Spain*  landete@umh.es

[3]*Departamento de Estatística e Investigação Operacional. Centro de Matemática, Aplicações Fundamentais e Investigação Operacional. Universidade de Lisboa, Portugal,* fsgama@fc.ul.pt

The capacitated facility location problem is a core problem in Location Science (see [3] for a survey of fixed-charge facility location problems). In this work we consider a variant of this problem in which facilities may *cooperate* in order to adapt their capacities to the demand of their customers. In particular, we consider the situation in which there may be capacity transfer between facilities.

The existence of a potential flow between facilities leads to a redefinition of the capacity of a facility, which results from the original one plus the amount received from other facilities minus the amount sent to other facilities. This redefinition may lead to multiple applications, see [1] and [2].

In this work we introduce this new variant and provide a mixed-integer linear formulation for it. Then, we strengthen the model by adding several families of valid inequalities. Some of them are similar to classical valid inequalities of the capacitated facility location problem, while some other are specific for this variant. Computational results illustrate the benefits of this variant.

# Distance-Constrained Close Enough Arc Routing Problem:
# Polyhedral study and B&C algorithm

Ángel Corberán[1], Isaac Plana[2], Miguel Reula[1], and José María Sanchis [3]

[1]*Dept. d'Estadística i Investigació Operativa, Universitat de València, Spain*

[2]*Dept. de Matemáticas para la Economía y la Empresa, Universitat de València, Spain*

[3]*Dept. de Matemática Aplicada, Universidad Politécnica de Valencia, Spain*

Recent technological advances allow many logistic problems to be solved more easily and with less cost. In particular, in meter reading problems, the companies can collect remotely the consumption data of their customers due to radio frequency technology (RFID). In some cases, a vehicle with a receiver travels through a neighborhood and, if it gets within a certain distance of a meter, the receiver is able to record the gas, water, or electricity consumption. Therefore, the vehicle does not need to traverse all the streets where there are meters but a subset of them that are close enough to all meters. In this application, the Close-Enough Arc Routing Problem (also known as Generalized Directed Rural Postman Problem) considers that each costumer is not located in a specific arc, but can be served whenever a vehicle traverses any arc of a given subset. We deal with a generalization of this problem, the Distance-Constrained Close Enough Arc Routing Problem (DC-CEARP) in which a fleet of vehicles is available. The vehicles have to leave from and return to the depot and the length of their routes must not exceed a maximum distance (or time). Several formulations and exact algorithms for this problem were proposed in Ávila et al. [1].

Here, we propose a new formulation for the DC-CEARP that combines two types of variables used in the formulations introduced in [1]. An exhaustive study of the associated polyhedron by describing several new families of valid inequalities is carried out. Moreover, a branch-and-cut in-corporating separation algorithms for the new valid inequalities and the upper bounds obtained with the matheuristic described in [2], has been implemented. Extensive computational experiments have been performed on a set of benchmark instances and the results are compared with those obtained with the algorithms proposed in Ávila et al. [1].

# Towards a stochastic programming modeling framework for districting

Antonio Diglio[1], Stefan Nickel[2,3], and Francisco Saldanha da Gama[4,5]

[1]*Università degli Studi di Napoli Federico II, Department of Industrial Engineering (DII), Piazzale Tecchio, 80 - 80125 Naples, Italy,*   antonio.diglio@unina.it

[2]*Institute of Operations Research, Karlsruhe Institute of Technology (KIT), Karlsruhe, Germany,*   stefan.nickel@kit.edu

[3]*Department of Logistics and Supply Chain Optimization, Research Center for Information Technology (FZI), Karlsruhe, Germany.*

[4]*Dept. Estatística e Investigação Operacional, Faculdade de Ciências da Universidade de Lisboa, 1749-016 Lisboa, Portugal,*   faconceicao@fc.ul.pt

[5]*Centro de Matemática, Aplicações Fundamentais e Investigação Operacional, Faculdade de Ciências da Universidade de Lisboa, 1749-016 Lisboa, Portugal.*

Districting Problems (DPs) aim at partitioning a set of basic geographic areas, named Territorial Units (TUs), into a set of larger clusters, called districts, according to some planning criteria. The latter typically refer to balancing, contiguity and compactness [1].

One aspect of practical relevance in DPs concerns the need to cope with changing demand. Depending on the particular problem we are dealing with, different possibilities emerge. One is to assume a reactive posture and solve a so-called redistricting problem [2].

One alternative to cope with demand changes is to embed time in the optimization models, when accurate forecasts for the demand are available [3]. Finally, if demand changes are uncertain then embedding uncertainty in the models may be desirable and, hence, a stochastic programming model emerges as appropriate.

In this paper, we introduce a Stochastic Districting Problem with Recourse (SDPR) whose aim is to partition a given set of TUs into a prefixed number of clusters in order to maximize the overall compactness and to meet balancing constraints, expressed in terms of average demand per district.

Demands associated to each TU are modeled as random variables. The problem is treated as two-stage stochastic program with recourse where we consider uncertainty to be captured by a finite set of scenarios. Districts are created in the first stage by allocating the basic areas to those TUs chosen as representative (centers) of the districts. In the second stage, i.e., after demand becomes known, balancing requirements are to be met. This is ensured by means of two recourse actions: outsourcing and reassignment of territory units. The latter consists of solving a redistricting problem. The objective function accounts for the total expected cost that includes the cost for the first stage territory design plus the expected cost incurred at the second stage by outsourcing and reassignment. The (re)assignment costs are associated with the distances between territory units which means that the focus is on the compactness of the solution.

The new modeling framework proposed is tested using four geographical instances built using real data from a province in northern Italy. In total, we generate 384 instances by diversifying the input parameters of the model. Extensions to the investigated problem triggered by features of practical relevance are also discussed. In particular, constraints accounting for the maximum territorial dispersion, the maximum allowed number of reallocations in the second stage and the similarity w.r.t. an existing plan are introduced. Preliminary computational experiments show how the incorporation of these features affect the properties of the solutions obtained.

In summary, the contribution of this paper to the literature is threefold: (i) to introduce a new modeling framework for a two-stage stochastic districting problem; (ii) to embed redistricting decisions as a way to hedge against uncertainty; (iii) to show the relevance of capturing stochasticity in districting problems.

# Solving Discrete Ordered Median Problems with Induced Order[*]

Enrique Domínguez[1] and Alfredo Marín[2]

[1]*Universidad de Málaga, Málaga, Spain,*  enriqued@lcc.uma.es

[2]*Universidad de Murcia, Murcia, Spain,*  amarin@um.es

Ordered median functions have been developed to model flexible discrete location problems. A weight is associated to the distance from a customer to its closest facility, depending on the position of that distance relative to the distances of all the customers. In this paper, the above idea is extended by adding a second type of facility and, consequently, a second weight, whose values are based on the position of the first weights. An integer programming formulation is provided in this work for solving this kind of models.

## 1.    Introduction

Discrete Location has a wide range of applications but is also an area that provides researchers with a rich variety of theoretical challenges. Discrete location problems involve a finite set of candidate sites where facilities can be installed, and a finite set of customers to be served from these facilities.

The increasing number of similar discrete location models made it necessary to develop new flexible location models. To that end, [5] dealt with an objective function that generalized the objective functions mentioned above. In order to accomplish this goal, a weight was applied to the allocation cost of each customer, depending on the position of that cost relative to the costs associated to the rest of customers. The inherent "sorting" aspect of the problem gave formulations and solutions a new dimension. The resulting problem, the so-called *Discrete Ordered Median Problem* (DOMP),

---

[*]Supported by the Ministerio de Economía y Competitividad, project MTM2015-65915-R

has received considerable attention. Many papers have been devoted to the development of more tricky formulations and also to the study of several variants. For instance, in [1], the first integer programming formulations that could solve instances of medium size were presented. A successful formulation based on a different paradigm to sort binary vectors inside an integer programming formulation was the key that allowed to solve larger instances, see [3] and the improved version in [4]. Several alternative formulations and variations have been recently compared in [2], and a recent survey on the DOMP and related problems is [6].

All the aforementioned models share the common idea of sorting distances or costs and then multiply them, in the given order, by a previously known constant.

In this paper, we propose a generalization of this idea through the modification of the second stage. After sorting the customers, the given constants will be multiplied, in the order determined in the first stage, by a different measure of the customer. An integer linear programming formulation is presented to modelize the proposed generalization and some experimental results are discussed.

# Minimum distance regulation and entry deterrence through location decisions

Javier Elizalde[1] and Ignacio Rodríguez-Carreño

[1] *University of Navarra, Campus Universitario, Edificio Amigos, 31009 Pamplona, Spain,* jelizalde@unav.es

This paper analyses the location strategies and the resulting market structure in a model of spatial competition, illustrating location in two dimensions, when there is a restriction of minimum distance between plants. This type of regulation exists in some retail markets, such a drugstores, aiming to avoid agglomeration and provide accessibility for all consumers, and may change the optimal location decisions of managers with the purpose of reducing the eventual number of competitors. The latter becomes endogenously determined by the size of the market and the distance rule and we evaluate the welfare consequences of the firms' location strategies when they take their decisions with the purpose of deterring additional entry.

We describe a theoretical model of spatial competition in two dimensions and solve for the equilibria through algorithmic simulations. The eventual number of active firms becomes endogenously determined by the size of the market and the distance rule. In a sequential entry game, we obtain a location equilibrium for a wide range of the binding distance rule and compare the equilibria reached under two types of firm behavior: with a simple maximum capture behavior and with entry deterrence strategies. We then discuss the results in terms of welfare in order to assess the effect of such regulatory policies and the distortions in firm's location decisions that they imply.

The main finding of the paper is that, with a minimum distance constraint, location equilibrium exists for each level of minimum distance which allows for two or more firms. Even though entry deterrence activities by incumbent firms tend to reduce the level of consumers' welfare as it tends to reduce the number of firms for some levels of minimum distance, it may in some cases be welfare enhancing as it may lead to a more even distribution of firms in the plane reducing the distance travelled by the average consumer.

# Upgraded network $p$-median problem [*]

Inmaculada Espejo[1] and Alfredo Marín[2]

[1]*Universidad de Cádiz*,　inmaculada.espejo@uca.es

[2]*Universidad de Murcia*,　amarin@um.es

A directed network is given. The allocation cost of a node $i$ to another one $j$ is given by the length of the shortest path from $i$ to $j$. At most $p$ nodes can be chosen as medians, and every node is allocated to the median of minimum cost (ties arbitrarily broken). Minimizing the sum of the allocation costs is known as *p-median problem on a network*. An optimal solution is represented in Figure 1 for 25 nodes (points on the plane) and $p = 3$ using Euclidean distances between their endpoints as costs of the edges.

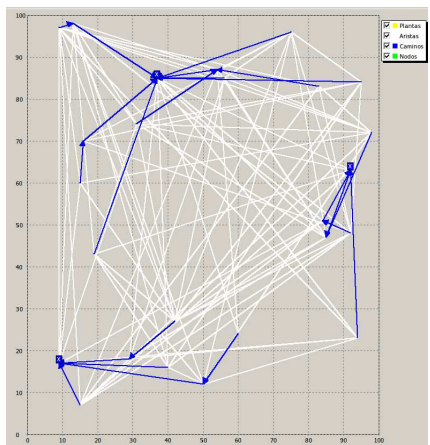

*Figure 1.*　　Optimal solution to a $p$-median problem on a network

[*]Supported by the Ministerio de Economía y Competitividad, project MTM2015-65915-R and Ministerio de Ciencia e Innovación, project MTM2016-74983-C2-2-R

The upgrading of a network can be done by either improving in some way its nodes (see e.g. [1]) or its arcs (see e.g. [2]). We consider upgrading of arcs, consisting in reducing their lenghts until a limit of $B$ units overall. This is done in a second stage, that is to say, the medians are chosen, then the nodes are allocated to the closest medians and finally the arcs to be upgraded and how much reduction is applied to each of them are decided. This is what we call the *Upgraded network p-median problem*.

We formulate the problem as an Integer Programming Problem, derive some properties of any optimal solution, develop valid inequalities and present preliminary computational results. Closest assignment constraints (see [3]) are required to avoid nodes to follow non-shortest paths that are more congested and therefore more likely to be upgraded, with a smaller total cost.

# On some classes of combinatorial optimization problems with non-convex neighborhoods

Inmaculada Espejo[1], Justo Puerto[2,3], and Antonio M. Rodríguez-Chía[4]

[1]*Departamento de Estadística e Investigación Operativa, Universidad de Cádiz, Spain,* inmaculada.espejo@uca.es

[2]*Departamento de Estadística e Investigación Operativa, Universidad de Sevilla, Spain* puerto@us.es

[3]*IMUS Instituto de Matemáticas, Universidad de Sevilla, Spain* puerto@us.es

[4]*Departamento de Estadística e Investigación Operativa, Universidad de Cádiz, Spain* antonio.rodriguezchia@uca.es

This paper provides formulations for some combinatorial optimization problems with neighborhoods which are not necessarily convex.

## 1.     Introduction

Combinatorial optimization has important applications in real world situations. Many such applications can be formulated as optimization problems defined on graphs, this is the case of planning shortest paths, spanning trees and matching among others. Recently, some extensions of these problems have been considered using convex regions, so-called neighborhoods, instead of vertices. In particular, the Traveling Salesman Problem for polygonal regions, discs and neighborhoods which are a set of disjoint convex full dimension objects was studied in [2,4,5]. The Minimum Spanning Tree Problem where neighborhoods are sets of disjoint discs and rectangles or second order cone representable was analyzed in [1,3,6].

In this paper, we deal with combinatorial optimization problems on graphs where vertices can be chosen in non-necessarily convex neighbor-hoods. We will develop mixed integer non linear programming formulations and study the structure of this kind of problems.

# A heuristic approach for the multi-drop Truck-Drone cooperative routing problem

Pedro L. González-R[1], David Canca[1], José L. Andrade-Pineda[2], Marcos Calle[1], and José M. León[1]

[1]*Department of Industrial Engineering and Management Science,School of Engineering, University of Seville, Seville. Spain,*  pedroluis@us.es

[2]*Robotics, Vision & Control Group, School of Engineering, University of Seville, Seville, Spain,*  jandrade@us.es

Recently there have been significant developments and applications in the field of unmanned aerial vehicles (UAVs). In few years, these applications will be fully integrated into our lives. The practical application and use of UAVs presents a number of problems that are of a different nature to the specific technology of the components involved. Among them, the most relevant problem deriving from the use of UAVs in logistics distribution tasks is the so-called "last mile" delivery. In the present work, we focus on the resolution of the truck-drone team logistics problem. The problems of tandem routing have a complex structure and have only been partially addressed in the scientific literature. The use of UAVs raises a series of restrictions and considerations that did not appear previously in routing problems; most notably, aspects such as the limited power-life of the batteries used by the UAVs and the definition of replacement or charging points. These difficulties have until now limited the mathematical formulation of truck-drone routing problems and their resolution to mainly small-size cases. To overcome these limitations we propose an iterated greedy heuristic based on the iterative process of destruction and reconstruction of solutions. This process is orchestrated by a global optimization scheme using a simulated annealing (SA) algorithm. We test our approach in a large set of instances of different sizes taken from literature. The obtained results are quite promising, even for large-size scenarios.

# 1. Introduction

Our research covers some of the gaps identified in the conceptualization of new working models around the use of drones in logistic scenarios. In particular, we focus on the applications where a drone trip is capable of serving several locations (every trip comprises of a launching, the visit to a typically small number of locations and a landing in the ground vehicle at the next rendezvous point). This differs from the commonly accepted hypothesis in last-mile delivery literature which in general considers that only one single location is visited at each drone's trip. In our approach to the problem, every location stands for a customer that can be served either by drone or by truck. We assume the truck stops at certain customer sites (which are not predefined) and hence, the drone is only allowed to merge with the ground vehicle at these locations. Once landed, the drone always gains a fully charged battery and would then be ready to start a new trip or stay at the truck while the truck carries it to a new service area.

We present a novel truck-drone team logistic (TDTL) mathematical formulation, allowing multi-drop routes for the drone. Under the assumption of infinite travel autonomy for the truck and a limited battery life for the drone, we plan the synchronization events –namely, the sites wherein the drone will gain a fully-charged new battery- with the objective of minimizing the makespan. We assume that the truck-drone team starts from an origin (depot) to serve a set of locations (customers), each one reachable by one of both vehicles, the truck or drone and, without pre-established routes (open) neither for the truck nor for drone.

In order to solve real-life instances of the above challenging mixed integer model (MIP), we propose a heuristic method with an innovative coding scheme, a fast and high-quality constructive two-step procedure for obtaining the initial solution and two original local search procedures for improving solutions. We code the solutions with a tuple formed by a node-sequence vector (defining the order of visits) and a resource-type vector containing the type of resource that visits each node (truck, drone or truck-drone) in the node-sequence vector.

The developed heuristic allows its transfer to real-life cases, since it is able to obtain solutions of relatively good quality for large-size problems in short computation time.

# An approach to obtain the Domatic Number of a graph by block decomposition

Mercedes Landete[1] and José Luis Sainz-Pardo[1]

[1]*Centro de Investigación Operativa,*
*Universidad Miguel Hernández de Elche, Spain,*
*landete@umh.es, jose.sainz-pardo@umh.es*

The domatic number problem is the NP-complete problem of partitioning a given graph into a maximum number of disjoint dominating sets. We propose an algorithm to obtain the domatic number of a graph by block decomposition. This method reduces the computational complexity with respect the state-of-the-art methods, and this reduction is quantified.

## 1.     Introduction

Obtaining the maximum number of disjoint dominating sets of a graph, this is the Domatic Number (DN), can be applied for example to answer the question of how many types of essential resources (hospitals, universities, computer servers, antennas, etc.) can be located on a network being them accessible for each node? Although the most effective ways to calculate it are based on algorithms, DN also can be formulated by the following linear integer model:

$$(DN(G)) \max \sum_{k=1}^{\delta+1} u_k \tag{1}$$

**s.t.**

$$x_u^k + \sum_{v:(u,v)\in E} x_v^k \geq u_k \quad u \in V, k \in K \tag{2}$$

$$\sum_{k=1}^{\delta+1} x_u^k \leq 1 \quad\quad\quad u \in V \tag{3}$$

$$u_k \in \{0,1\}, x_u^k \in \{0,1\} \quad u \in V, k \in K \tag{4}$$

where $u_k$ takes value $1$ if there exists some vertex belonging to the $k$ disjoint dominating set. Constraints (2) express domination, constraints (3) express disjointness among sets and (4) express variables are binary. Finally, objective function (1) maximizes the number of disjoint dominating sets.

# 2.     Computational complexity

The computational complexity of $DN(G)$ obtained by van Rooij Algotihm ( [2]), which is based on inclusion-exclusion technique, is $\mathcal{O}(2.7139^n)$ where $n = |V|$. For each block $B$, let $m_B = n_B + |N[B]| - |N[B] \cap B|$, $n_B$ its number of nodes, and $N[B]$ its closed neighbourhood. Then the computational complexity of $DN(G)$ by block decomposition is $\mathcal{O}(2.7139^M)$ where $M = max\{m_B, B \in blocks(G)\}$.

# Promoting the selective collection of urban solid waste by means an optimal deployment of routes for mobile eco-points [*]

Juan A. Mesa[1], Francisco A. Ortega[1], and Ramón Piedra-de-la-Cuadra[1]

[1]*IMUS, Universidad de Sevilla, Seville, Spain,* jmesa@us.es riejos@us.es rpiedra@us.es

The concept of *municipal waste* is referred in Eurostat (2012) as those waste which is mainly produced by households and other sources such as commerce, offices and public institutions that can generate similar wastes to the former case.

The composition of municipal urban waste varies depending on the standard of living of the population, the economic activity developed by their inhabitants and the climate of the region. Depending on these factors, certain products will eventually be more used and, subsequently, the corresponding waste will be generated. According to the National Urban Waste Plan (PNRU) 2000-2006, the average distribution with respect to the total weight in Spain of the main components of urban waste is as shown below:

- Organic matter (44.06%), derived from food scraps or activities related to gardening.

- Paper and cardboard (21.18%), including newspapers, boxes or containers where paper and cardboard are present.

- Plastic (10.59%), material which is used in almost all industrial sectors and for the manufacture of a wide range of products and for the general packaging.

- Glass (6.93%), mainly empty bottles.

[*]This research has been partially supported by the Spanish Ministry of Economy and Competitiveness through grant MTM2015-67706-P (MINECO/FEDER, UE). This support is gratefully acknowledged.

Moreover, Ferric and non-ferrous metals (4.11%), like tinplate and aluminum, used as a material for the production of carbonated beverage cans and tetra-brik, Woods (0.96 %), usually presented in the form of furniture, and Others (representing 12.17%).

This group has a very varied composition and due to the nature of some of the elements that compose it requires special attention, since some can be considered as hazardous waste. The interest of this article focuses on this last heterogeneous group of diverse components that are not as frequent in human consumption as to justify the permanent location of a container in the street, as it happens with organic waste, glass, plastic bottles and paper or cardboard.

In Spain, the so-called *eco-points* are large waste containers with watertight sections to separate the collection of a wide variety of items: small appliances, electronic scrap, mobile, small batteries, chargers, or books (for the exchange between citizens), toner and cartridges of ink, used vegetable oil, aluminum and plastic coffee capsules, CDs, DVDs and video and audio tapes, as well as small tires, batteries, glass (mirrors, windows, glass), syringes and needles, used low-energy lamps, mercury thermometers or radiographs and photographic material. The eco-point configuration is related to the bin packing problem, where items (bins) of different volumes must be packed into a finite number of containers, each of a fixed given volume in a way that minimizes the number of containers used.

The so-called *mobile eco-points* follow an established route, visiting all the neighborhoods of the city in an itinerant way. The service is provided at stops identified on public roads during a temporary period that has previously been made known to neighbors. The container is placed, for example, on a Monday and, the following Monday, it is moved to a new point in the city.

It is about separating small format waste so that it can be recycled with the collaboration of citizens. In addition, it is possible, with the presence of the eco-point in the street, to sensitize citizens about the need to be involved in recycling and advance in the control of waste in the city.

In this work an optimization model has been formulated for the deployment of routes for mobile eco-points for the selective collection of urban solid waste. An algorithm has also been developed to solve the proposed model of mathematical programming. The performance evaluation of the developed methodology has been carried out through a computational experience in an area belonging to the area of Seville (Spain).

# A Facility Location Problem with Transhipment points

Alejandro Moya-Martínez[1], Mercedes Landete[1], and Juan Francisco Monge[1]

[1]*Centro de Investigación Operativa,*
*Universidad Miguel Hernández de Elche, Spain,*
a.moya@umh.es, monge@umh.es, landete@umh.es

## 1.     Introduction

The $p - median$ is one of the basic models in discrete location theory and it is classified as NP-hard. The aim of the problem is to locate p facilities in such a way that the transport costs is minimized. In this paper, we propose the problem of opening facilities and transhipment points so that any customer is served by an open facility or by a close-enough open transhipment point. The main feature in our model compared with the classical p-median problem lies in the fact that any customer can move to a close-enough transhipment point for satisfying his demand. In doing so, we reduce the company costs since the distance is reduced. Different models and computational results will be presented in this work.

# An application of the $p$-median problem in optimal portfolio selection[*]

Justo Puerto[1], Moisés Rodríguez-Madrena[1], and Andrea Scozzari[2]

[1]*IMUS, Universidad de Sevilla, Sevilla, Spain,*   puerto@us.es, madrena@us.es

[2]*Faculty of Economics, Università degli Studi Niccolò Cusano Roma, Italy,*
andrea.scozzari@unicusano.it

In this work we propose a novel framework for portfolio selection that combines the specific features of a clustering and a portfolio optimization techniques through the global solution of a hard Mixed-Integer Linear Programming problem. In particular, we endow the assets network with a metric based on correlation coefficients between assets' returns, and show how classical location problems on networks can be embedded in the portfolio optimization models in order to cluster assets.

## 1.    The problem

Given a set of assets and an investment capital of a stilized investor, the Portfolio Optimization Problem consists on determining the amount of the capital to be invested in each asset in order to build the most profitable portfolio. The Portfolio Optimization Problem is classically modeled as a mean-risk bi-criteria optimization problem (see the seminal paper of Markowitz in 1952 [4]):

$$\max\{[\mu(\mathbf{x}), -\varrho(\mathbf{x})] : \quad \mathbf{x} \in \Delta\},$$

where the mean rate of return $\mu(\mathbf{x})$ of the portfolio is maximized and a risk measure $\varrho(\mathbf{x})$ is minimized, being $\Delta$ the set of feasible portfolios.

[*]The first and the second authors have been partially supported by MINECO Spanish/FEDER grant number MTM2016-74983-C02-01.

New mathematical programming models and techniques are still needed in order to efficiently solve the Portfolio Optimization Problem. A relatively recent promising line of research is to exploit clustering information of an assets network in order to develop new portfolio optimization paradigms [1, 5]. In this work we endow the assets network with a metric [3] and show how classical location problems on networks can be used for asset clustering (for the interpretation of the $p$-median problem in terms of cluster analysis the reader if referred to [2]). In particular, we add a new criterion to the Portfolio Optimization Problem which measures, by means of the objective function $F_p(\mathbf{x})$ of the $p$-median problem, the degree of representation of the selected assets with respect to the non-selected ones in a portfolio $\mathbf{x}$ with exactly $p$ assets:

$$\max\{[\mu(\mathbf{x}), -\varrho(\mathbf{x}), -F_p(\mathbf{x})] : \quad \mathbf{x} \in \Delta\}.$$

# 2. Solution approach and results

We propose a Mixed-Integer Linear Programming formulation for dealing with this problem. The usefulness of our approach is validated reporting some computational experiments: our model was tested on real financial datasets, compared to some benchmark models, and found to give good results in terms of realized profit.

# A matheuristic to solve the MDVRP with Vehicle Interchanges[*]

Victoria Rebillas-Loredo[1], Maria Albareda-Sambola[1], Juan A. Díaz[2], and Dolores E. Luna-Reyes[2]

[1]*Universitat Politècnica de Catalunya, Cataluña, España,*  victoria.rebillas@upc.edu, maria.albareda@upc.edu

[2]*Universidad de las Américas Puebla, Cholula, Puebla, México,*  juana.diaz@udlap.mx, dolorese.luna@udlap.mx

This work defines a matheuristic for the Multi-Depot Vehicle Routing Problem with Vehicle Interchanges. This is a node-routing problem with capacitated vehicles and route duration limits for the drivers [1]. In order to try to exploit as much as possible these two limited resources, it is allowed that, at predefined interchange locations, two drivers meet and interchange their vehicles. With this strategy, it is possible to reduce the total costs and the number of used routes with respect to the classical approach: the Multi-Depot Vehicle Routing Problem. It should be noted that the Multi-Depot Vehicle Routing Problem is more challenging and sophisticated than the single-depot Vehicle Routing Problem. From the complexity point of view, the Multi-Depot Vehicle Routing Problem with Vehicle Interchanges is NP-Hard, since it is an extension of the classical problem, which is already NP-Hard. For this reason, we decide to implement a matheuristic.

For the heuristic algorithm a set of "partial routes" are generated with the aim of building feasible routes for both, classical and the Multi-Depot Vehicle Routing Problem with Vehicle Interchanges problems. These routes are generated with a GRASP, which uses a utility function that combines the cost and the vehicle capacity usage. Then to improve the quality of these partial routes, a two-opt is used. These partial routes are then combined in the best possible way by solving a mathematical program. Finally,

[*]Thanks to Spanish Ministry of Economy and Competitiveness end EDRF funds through grant MTM2015-63779-R (MINECO/FEDER)

Local Search is applied to the final solution. Computational results on a series of instances show the capacity of the method to produce high quality solutions.

**Keywords:** combinatorial optimization, routing, the Multi-Depot Vehicle Routing Problem, Rich Vehicle Routing Problem, vehicle interchanges, heuristic algorithm, local-search.

# Orienteering with synchronization in a telescope scheduling problem

Jorge Riera-Ledesma[1], Juan-José Salazar-González[2], and Francisco Garzón[3]

[1]*Departamento de Ingeniería Informática y de Sistemas, Universidad de La Laguna*   jriera@ull.es

[2]*Departamento de Matemáticas, Estadística e Investigación Operativa, Universidad de La Laguna*   jjsalaza@ull.es

[3]*Departamento de Astrofísica, Universidad de La Laguna*   fgarzon@ull.es

We introduce a new optimization problem arising in the management of an observation instrument for a telescope. It consists of a set of configurable devices which allows the simultaneous study of various astronomic objects. The observation of an object may require the synchronized configuration of several devices. Astronomers using this instrument propose a list of objects to observe. Because the time window assigned to each proposal is quite limited, the astronomers associate a priority with each object. This paper describes and solves the problem of selecting the objects to observe. The problem also determines an optimal sequence of configurations for each device to maximize the total priority of the selection, subject to some synchronization issues and a time limit to conclude the last observation. We show three mathematical formulations, where one of them being a set-partitioning model. Its master problem manages the synchronization constraints using combinatorial cuts, while its subproblem performs the time limitation constraints using a column generation algorithm. We propose two branch-and-price-and-cut algorithms for solving the set-partitioning model. The paper discusses an extensive computational experience showing the performance of the algorithms. The algorithms solve to optimality instances involving up to one-hundred and fifty objects under a low level of synchronization, and up to sixty objects under a high level of synchronization.

# Network Design under Uncertain Demand: an Alternative Capacitated Location Decision Framework

Diego Ruiz-Hernández[1], Mozart M.B.C Menezes[2], and Oihab Allal-Cherif[3]

[1]*Sheffield University Management School, Conduit Road, Sheffield, UK,*
d.ruiz-hernandez@shefield.ac.uk

[2]*NEOMA Business School, 9 Rue d'Athénes, Paris, France*
mozart.menezes@neoma-bs.fr

[3]*NEOMA Business School, 9 Rue d'Athénes, Paris, France*
oihab.allal-cherif@neoma-bs.fr

In this work we address the problem of designing a distribution network for a new product (or, alternatively, for the release of a new product in a new market). The complexity of this problem becomes magnified because the location and capacity decisions are made long before the product is released to the market and, thus, knowledge about demand is limited. Moreover, in the short term, location and capacity are one-shot decisions. An example of an important area where this kind of problems may appear is the one of pre-positioning of temporary facilities for humanitarian aid, where adapting existing buildings or deploying new facilities for housing refugee centres is a one go decision that cannot be revisited. In such cases, the need for the facility is in general short lived, while the real demand is highly uncertain. The relevance of location and capacity decisions is better appreciated by considering that about 80% of the supply chain costs are locked-in when facilities' location and capacity are fixed.

The problem is formulated as a Newsvendor model integrated in a p-Median facility location framework. Additionally, we allow production to be either fully manufactured in-house, or partially outsourced. This confers our model important flexibility with respect to capacity decisions. We propose a heuristic for solving the simultaneous location/capacity problem and compare the obtained solution against the one obtained when the problem is solved by the independent use of a p-Median formulation and a Newsvendor approach for capacity determination. Although the deterministic variant of the capacitated facility location problem has been thoroughly addressed in lilterature, and demand stochasticity has also been included in facility location frameworks, to our knowledge this is the first time that the problem of simultaneously locating facilities and determining their capacities is addressed with explicit consideration of the demand's variability. We show that for the single facility case, the expected profit of the strategic problem is non-decreasing and concave in capacity, resulting in a uniquely determined optimal capacity. We further show that when the facility's location is fixed, the problem of choosing the capacity becomes a variation of the classical Newsvendor model. A critical-ratio-based heuristic is proposed for the multi-facility case. An illustrative example is provided.

# A Location-Allocation Model for Bio-Waste Management

Dolores R. Santos-Peñate[1], Rafael R. Suárez-Vega[1], and Carmen Florido de la Nuez[2]

[1]*Dpto de Métodos Cuantitativos en Economía y Gestión, and TIDES, Universidad de Las Palmas de Gran Canaria, Spain,*  {dr.santos, rafael.suarez}@ulpgc.es

[2]*Dpto de Análisis Económico Aplicado , Universidad de Las Palmas de Gran Canaria, Spain,*  carmen.florido@ulpgc.es

The tourism sector generates a large amount of bio-waste from the activity developed in kitchens and restaurants (food waste), and gardens. In some cases this bio-waste is transferred to a treatment facility belonging to the regional government where the waste is transformed into compost or some other product. However, the profitability of the process is usually very low (in some real cases only around 8% of the bio-waste is treated) and most of the waste ends up in a landfill. In accordance with sustainability and environmental respect practices, some hotels have incorporated waste separation processes and composting equipments managed by the hotel staff. The compost produced is used in the hotels or in the agricultural sector. On the other hand, forest residues, and those generated in parks and gardens, can be used to produce pellets. Pellets are utilized as combustible material to generate energy and also as absorbent and balancing material in the biological process carried out in the composting machines.

In this work we consider a hotel chain that wishes to install composting equipment to treat the bio-waste generated and produce compost, taking into account that the waste from its gardens can be used to produce compost and pellets. These private facilities coexist with public facilities, so that untreated waste in hotels would have to be transferred to public facilities (scheme in Figure 1). The problem consists in determining where the treatment facilities (composting machines) should be opened and how the generated waste and the product (compost and pellets) should be allocated to the treatment and demand points, respectively, so that certain objectives are achieved. To solve the problem we formulate a multi-objective MILP model. The aim is to minimize costs subject to certain capacity and demand constraints. An example where the problem is solved using an exact procedure is presented.

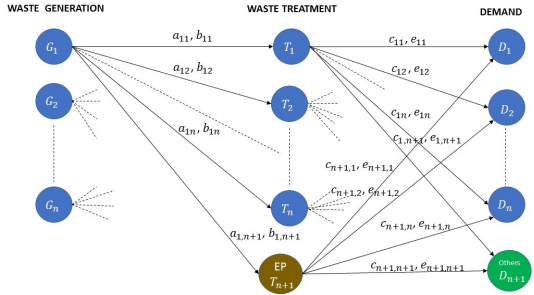

*Figure 1.* Scenario for two types of bio-waste and two products

# XPP with Neighborhoods

Carlos Valverde[1] and Justo Puerto[2]

[1]*University of Seville, Seville, Spain,*  carvalmar4@us.es

[2]*University of Seville, Seville, Spain,*  puerto@us.es

This paper presents a routing problem that combines elements from the crossing postman and the travelling salesman problems. The elements to be toured are general neighborhoods induced by $l_p$-norms and line segments. We assume that the considered vehicle has to travel a prefixed percentage of each line segment and that must visit each neighborhood only once. This problem has natural applications in the case of drone routing assuming that the vehicles can access and leave the service areas at any point and not only from prespecified points as the vertices of the edges.

We provide several formulations for the problem and a computational comparison among them. We also introduce some heuristic algorithms that generate reasonably good initial solutions that help in solving the original formulations.

# Author Index

# X Worksh

# Locat

# Analysis

# Proble

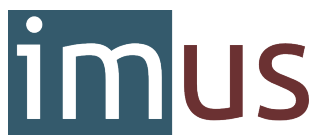
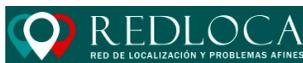
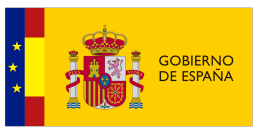
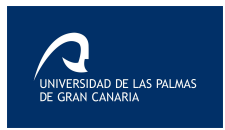
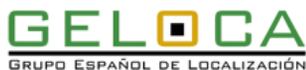
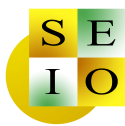
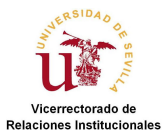
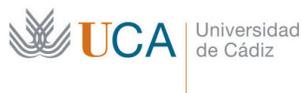